\documentclass[final,5p,times,twocolumn]{elsarticle}
\usepackage{amsmath}
\usepackage{amssymb}
\journal{Optik}

\begin{document}
\begin{frontmatter}
\title{Is the Abraham electromagnetic force physical?}

\author{Changbiao Wang\corref{cor1}}
\cortext[cor1]{Tel: +1 203 893 2084}
\ead{changbiao\_wang@yahoo.com}
\address{ShangGang Group, 70 Huntington Road, Apartment 11, New Haven, CT 06512, USA}

\begin{abstract}
A conventional general electromagnetic force definition has been widely used to analyze radiation forces in dielectric media in published research works.  However in this paper, we would like to indicate that this conventional force definition is flawed.
\end{abstract}
\end{frontmatter}

In a recent paper \cite{r1}, Brevik and Ellingsen proposed a novel idea to experimentally measure the time-dependent Abraham force which is a component of the conventional general electromagnetic (EM) force definition \cite{r2}.  However in this paper, we would like to indicate that this general EM force definition itself is flawed.

The conventional general EM force definition, namely the expression of the force exerted by an EM field on a unit volume of isotropic dielectric material, is given by \cite{r2}
\begin{equation}
\mathbf{f}=\mathbf{f}^{\mathrm{AM}}+\mathbf{f}^{\mathrm{A}},
\label{eq1}
\end{equation}
where $\mathbf{f}^{\mathrm{AM}}$ is the Abraham-Minkowski (term) force, and $\mathbf{f}^{\mathrm{A}}$ is the Abraham (term) force.\footnote
{As indicated by Stratton, if there is any deformation in the dielectric material, Eq.\ (\ref{eq1}) is ``manifestly incorrect'' because it does not include the forces associated with the material deformation \cite{r2}.  However Eq.\ (\ref{eq1}) is ``correct'' for the widely-used physical model of ``uniform'' isotropic medium, such as the one used in the work by Brevik and Ellingsen \cite{r1}, because the ``uniform'' material is ``rigid'' and no deformation is involved.  In other words, in the Stratton's sense, Eq.\ (\ref{eq1}) is the general EM force definition in a ``rigid'' isotropic medium.
}
  Note: For a uniform medium, Eq. (\ref{eq1}) is the same as the one given in \cite{r3}.  For a non-magnetic medium (relative permeability $\mu_r=1$) \emph{without any sources}, $\mathbf{f}^{\mathrm{AM}}$  and $\mathbf{f}^{\mathrm{A}}$ are, as shown by Brevik and Ellingsen \cite{r1,r4}, given by
\begin{equation}
\mathbf{f}^{\mathrm{AM}}=-\frac{1}{2}\epsilon_{0}\mathbf{E}^2\nabla n_d^2, 
\label{eq2}
\end{equation}
\begin{equation}
\mathbf{f}^{\mathrm{A}}=\frac{n_d^2-1}{c^2}\frac{\partial}{\partial t}(\mathbf{E}\times\mathbf{H}),
\label{eq3}
\end{equation}
where $\epsilon_0$ is the vacuum permittivity constant, $n_d=(\mu_r \epsilon_r)^{1/2}$ is the refractive index with $\epsilon_r$  the relative permittivity, and $c$ is the vacuum light speed.

In the idea proposed by Brevik and Ellingsen, the dielectric medium is assumed to be uniform ($\nabla n_d=0$), and thus $\mathbf{f}^{\mathrm{AM}}=0$ holds; $\mathbf{f}=\mathbf{f}^{\mathrm{AM}}+\mathbf{f}^{\mathrm{A}}=\mathbf{f}^{\mathrm{A}}$ is thought to form possibly measurable Abraham radiation torque in a micrometer-sized dielectric spherical resonator operating at the whispering gallery mode \cite{r1}.  Unfortunately, as shown below, this general EM force definition $\mathbf{f}=\mathbf{f}^{\mathrm{AM}}+\mathbf{f}^{\mathrm{A}}$ itself is flawed.

In principle, the correctness of $\mathbf{f}=\mathbf{f}^{\mathrm{AM}}+\mathbf{f}^{\mathrm{A}}$ as a general EM force definition cannot be legitimately affirmed by enumerating specific examples, no matter how many; however, the correctness can be directly negated by finding specific examples, even only one.  In the following, such a specific example is given to show why the conventional EM force definition $\mathbf{f}=\mathbf{f}^{\mathrm{AM}}+\mathbf{f}^{\mathrm{A}}$ is flawed.

An electromagnetic plane wave, although not practical, is a simplest strict solution of Maxwell equations, and it is often used to explore most fundamental physics.  For example, Einstein used a plane wave to develop his special theory of relativity and derived the well-known relativistic Doppler formula in free space \cite{r5}.  Thus if the general force definition $\mathbf{f}=\mathbf{f}^{\mathrm{AM}}+\mathbf{f}^{\mathrm{A}}$  is correct, it must withstand the test of a monochromatic plane wave in a non-dispersive, lossless, non-conducting, isotropic uniform medium. 

Suppose that the EM fields are given by $(\mathbf{E},\mathbf{B},\mathbf{D},\mathbf{H})=(\mathbf{E}_0,\mathbf{B}_0,\mathbf{D}_0,\mathbf{H}_0)\cos\Psi$ for the plane wave, where $\mathbf{E}_0$, $\mathbf{B}_0$, $\mathbf{D}_0$, and $\mathbf{H}_0$ are the constant amplitude vectors, and $\Psi=\omega t-\mathbf{k}_w\cdot\mathbf{x}$ is the phase function with $\omega$ the frequency and $\mathbf{k}_w$ the wave vector.  Since the medium is uniform,  $\mathbf{f}^{\mathrm{AM}}=0$ holds, and the general EM force $\mathbf{f}=\mathbf{f}^{\mathrm{AM}}+\mathbf{f}^{\mathrm{A}}$ is reduced to 
\begin{equation}
\mathbf{f}=\mathbf{f}^{\mathrm{A}}=\frac{n_d^2-1}{c^2}\frac{\partial}{\partial t}(\mathbf{E}\times\mathbf{H}).
\label{eq4}
\end{equation}
From Maxwell equations, the momentum conservation equation is given by \\
\begin{equation}
\frac{\partial}{\partial t}\left(\frac{\mathbf{E}\times\mathbf{H}}{c^2}\right)=-\nabla\cdot \check{\mathbf{T}}_A,
\label{eq5}
\end{equation} \\
where the stress tensor $\check{\mathbf{T}}_A$ is given by \\
\begin{equation}
\check{\mathbf{T}}_A=\boldsymbol{\beta}_{ph}^2[-(\mathbf{ED}+\mathbf{HB})+\check{\mathbf{I}}\frac{1}{2}(\mathbf{D}\cdot\mathbf{E}+\mathbf{B}\cdot\mathbf{H})],
\label{eq6}
\end{equation} \\
with $|\boldsymbol{\beta}_{ph}|=1/n_d$  the absolute phase velocity normalized to the light speed $c$, and $\check{\mathbf{I}}$  the unit tensor. 

Inserting $\mathbf{E}\times\mathbf{H}=\mathbf{E}_0\times\mathbf{H}_0\cos^2\Psi$ into Eq.\ (\ref{eq4}), we indeed have $\mathbf{f}=\mathbf{f}^{\mathrm{A}}\ne 0$ holding (except for those discrete points); thus $\mathbf{f}\ne 0$  looks like a ``force'', but that is not true.  This can be seen from the following analysis.

Inserting Eq.\ (\ref{eq5}) into Eq.\ (\ref{eq4}) we have $\mathbf{f}=-(n_d^2-1)\nabla\cdot\check{\mathbf{T}}_A\ne 0$.  From Eq.\ (\ref{eq6}), we know that $\check{\mathbf{T}}_A\varpropto\cos^2\Psi$ is a ``pure travelling-wave'' stress tensor, and thus the Abraham momentums flowing into and out from a differential box are usually different, resulting in $\nabla\cdot\check{\mathbf{T}}_A\ne 0 \Rightarrow \mathbf{f}\ne 0$. From this we can see that $\mathbf{f}\ne 0$  is resulting from the attribution of the ``pure travelling wave'' of tensor $\check{\mathbf{T}}_A$. This ``pure travelling-wave'' attribute will not produce any ``force effect'' on the medium.\footnote
{
That $\nabla\cdot\check{\mathbf{T}}_A\ne 0$  is an attribution of ``pure travelling wave'' of tensor $\check{\mathbf{T}}_A$  can be better understood from the free-space case where there is no dielectric medium but $\nabla\cdot\check{\mathbf{T}}_A\ne 0$  holds.
}$^{,}$\footnote
{
One might argue that to calculate the total force on a material \emph{differential} box, the bound \emph{surface} charge on the box boundaries needs to be considered \---- ``even if it is embedded within the surrounding material''.  However according to the uniform-medium model used in the paper, no sources are assumed and the Maxwell equation $\nabla\cdot\mathbf{D}=\epsilon_0 \epsilon_r\nabla\cdot\mathbf{E}=0$  holds, where $\mathbf{D}=\epsilon_0\mathbf{E}+\mathbf{P}$  with $\mathbf{P}$  the electric polarization, resulting in  $\nabla\cdot\mathbf{P}=0$, namely no polarized charge or bound surface charge on the \emph{differential} box boundaries, and thus there is no additional surface-charge-caused force.
}  
This phenomenon can be clearly understood through Einstein's light-quantum hypothesis: photons are the carriers of light momentum and energy.  Since the dielectric medium is assumed to be a non-dispersive, lossless, isotropic uniform medium, all the photons move \emph{uniformly} at the dielectric light speed $c/n_d$, and they do not have any momentum exchanges with the medium.  

Since $\mathbf{f}=\mathbf{f}^{\mathrm{AM}}+\mathbf{f}^{\mathrm{A}}=\mathbf{f}^{\mathrm{A}}\ne 0$ does not represent a force for a plane wave, $\mathbf{f}=\mathbf{f}^{\mathrm{AM}}+\mathbf{f}^{\mathrm{A}}$  cannot pass the plane-wave test, and $\mathbf{f}=\mathbf{f}^{\mathrm{AM}}+\mathbf{f}^{\mathrm{A}}$, as a general EM force definition, is flawed.  Unfortunately, this flawed EM force definition is widely accepted in the community \cite{r1,r4,r6,r7,r8}, and it is argued that the  $\mathbf{f}^{\mathrm{A}}$-term ``simply fluctuates out when averaged over an optical period in a stationary beam'', but ``it is in principle measurable'' \cite{r6}. 

In summary, we have shown that the conventional general EM force definition $\mathbf{f}=\mathbf{f}^{\mathrm{AM}}+\mathbf{f}^{\mathrm{A}}$  is flawed.  Specifically speaking, the Abraham term  $\mathbf{f}^{\mathrm{A}}=(n_d^2-1)(\partial/\partial t)(\mathbf{E}\times\mathbf{H})/c^2$ is not a ``physical EM force'' at all for a plane wave.  

It is interesting to point out that the conclusion obtained in the present paper is completely consistent with that obtained from analysis of dielectric Einstein-box thought experiment, which states that ``the momentum transfer from the light pulse to the box only takes place on the vacuum-medium interface, while the pulse edges located inside the uniform medium do not have any contributions to momentum transfer'' \cite{r9}.  This conclusion is also consistent with the recent assertion by Ramos and coworkers that ``there is no reason to assume the existence of'' the Abraham-term force \cite{r10}.

\appendix
\section{Abraham momentum conservation equation}
\label{A}
Isotropic medium is a special case of anisotropic media.  Below we will show that Eq.\ (\ref{eq5}) is also valid for an anisotropic medium. 

\textbf{Mathematical statement.} If a plane wave propagates in a lossless, non-dispersive, non-conducting, uniform anisotropic medium, the Abraham momentum conservation equation can be written as \cite{r11}\footnote
{
One might argue that Eq.\ (\ref{eq5}) or Eq.\ (\ref{eqA1}) is incorrect, and it should be replaced by $-\nabla\cdot\check{\mathbf{T}}_d-(\partial/\partial t)[(\mathbf{E}\times\mathbf{H})/c^2]=(n_d^2-1)[(\mathbf{E}\times\mathbf{H})/c^2]$  according to the review paper by Brevik \cite{r12}.  However this is apparently wrong, because  $(\partial/\partial t)[(\mathbf{E}\times\mathbf{H})/c^2]$ and $(n_d^2-1)[(\mathbf{E}\times\mathbf{H})/c^2]$ have different physical dimensions.  As we know, 20 kilograms + 5 meters usually does not produce any physical meanings, because kilogram and meter are of different physical dimensions.
} \\
\begin{equation}
\frac{\partial}{\partial t}\left(\frac{\mathbf{E}\times\mathbf{H}}{c^2}\right)=-\nabla\cdot \check{\mathbf{T}}_A,
\label{eqA1}
\end{equation} \\
where the Abraham stress tensor is given by \\
\begin{equation}
\check{\mathbf{T}}_A=\boldsymbol{\beta}_{ph}^2\left[-(\mathbf{ED}+\mathbf{HB})+\check{\mathbf{I}}\frac{1}{2}(\mathbf{D}\cdot\mathbf{E}+\mathbf{B}\cdot\mathbf{H})\right].
\label{eqA2}
\end{equation} \\
\indent In Abraham theory, $(\mathbf{E}\times\mathbf{H})/c^2$ is defined as EM momentum density.  Thus Eq.\ (\ref{eqA1}) means that in unit time, the total EM momentum flowing into a differential box is equal to the increase of the EM momentum in the box.

\textbf{Proof.}  For a monochromatic plane wave with a phase function of $\Psi=\omega t-\mathbf{k}_w\cdot\mathbf{x}$, Maxwell equations are simplified into  \\
\begin{equation}
\omega\mathbf{B}=\mathbf{k}_w\times\mathbf{E}, \hspace{2mm} \omega\mathbf{D}=-\mathbf{k}_w\times\mathbf{H},
\label{eqA3}
\end{equation} 
\begin{equation}
\mathbf{k}_w\cdot\mathbf{B}=0, \hspace{2mm} \mathbf{k}_w\cdot\mathbf{D}=0,
\label{eqA4}
\end{equation} \\
where the EM fields are given by $(\mathbf{E},\mathbf{B},\mathbf{D},\mathbf{H})=(\mathbf{E}_0,\mathbf{B}_0,\mathbf{D}_0,\mathbf{H}_0)\cos\Psi$, with $\mathbf{E}_0$, $\mathbf{B}_0$, $\mathbf{D}_0$, and $\mathbf{H}_0$ the real constant vectors.  The frequency $\omega$  and the wave vector $\mathbf{k}_w$ are real because the medium is assumed to be non-conducting and lossless.  

From Eq.\ (\ref{eqA3}) we have \\
\begin{equation}
\mathbf{D}\times\mathbf{B}=\left(\frac{\mathbf{D}\cdot\mathbf{E}}{\omega}\right)\mathbf{k}_w.
\label{eqA5}
\end{equation} 

By making cross products of $\mathbf{k}_w\times(\omega\mathbf{B}=\mathbf{k}_w\times\mathbf{E})$ and $\mathbf{k}_w\times(\omega\mathbf{D}=-\mathbf{k}_w\times\mathbf{H})$ from Eq. (\ref{eqA3}), with vector identity 
$\mathbf{a}\times(\mathbf{b}\times\mathbf{c})=(\mathbf{a}\cdot\mathbf{c})\mathbf{b}-(\mathbf{a}\cdot\mathbf{b})\mathbf{c}$ taken into account, we have 
\begin{equation}
\mathbf{E}=(\textbf{\^n}\cdot\mathbf{E})\textbf{\^n}-\mathbf{v}_{ph}\times\mathbf{B},
\label{eqA6}
\end{equation}
\begin{equation}
\mathbf{H}=(\textbf{\^n}\cdot\mathbf{H})\textbf{\^n}+\mathbf{v}_{ph}\times\mathbf{D},
\label{eqA7}
\end{equation}
where $\textbf{\^n}=\mathbf{k}_w/|\mathbf{k}_w|$ is the unit wave vector, and $\mathbf{v}_{ph}=\textbf{\^n}(\omega/|\mathbf{k}_w|)$ is the phase velocity.  The refractive index for anisotropic media is defined as $n_{d}=|\mathbf{k}_w|/|\omega/c|$, with $c$ the vacuum light speed, and thus the phase velocity also can be written as $\mathbf{v}_{ph}=\textbf{\^n}(\omega/|\omega|)(c/n_d)$.

By making inner products of $\textbf{H}\cdot(\omega\textbf{B}=\textbf{k}_w\times\textbf{E})$  and $\textbf{E}\cdot(\omega\textbf{D}=-\textbf{k}_w\times\textbf{H})$ from Eq. (\ref{eqA3}), with $\textbf{H}\cdot(\textbf{k}_w\times\textbf{E}) = \textbf{E}\cdot(-\textbf{k}_w\times\textbf{H})$ taken into account we obtain $\textbf{E}\cdot\textbf{D}=\textbf{B}\cdot\textbf{H}$. 

From Eqs. (\ref{eqA6}) and (\ref{eqA7}), and Eq.\ (\ref{eqA5}) we obtain 
\begin{align}
&\frac{\mathbf{E}\times\mathbf{H}}{c^2}=\frac{\boldsymbol{\beta}^2_{ph}}{\omega}\cos^2\Psi  \nonumber \\ 
&~~~\times[-(\mathbf{k}_w\cdot\mathbf{E}_0)\mathbf{D}_0
-(\mathbf{k}_w\cdot\mathbf{H}_0)\mathbf{B}_0+(\mathbf{D}_0\cdot\mathbf{E}_0)\mathbf{k}_w],
\label{eqA8}
\end{align}  
where $\boldsymbol{\beta}_{ph}=\mathbf{v}_{ph}/c$ is the normalized phase velocity.

With the help of $\nabla\cdot(\mathbf{ab})=(\nabla\cdot\mathbf{a})\mathbf{b}+\mathbf{a}\cdot(\nabla\mathbf{b})$ we obtain 
\begin{align}
\nabla\cdot[-(\mathbf{ED}&+\mathbf{HB})]=2\cos\Psi\sin\Psi  \nonumber 
\\ 
&~~~\times[-(\mathbf{k}_w\cdot\mathbf{E}_0)\mathbf{D}_0
-(\mathbf{k}_w\cdot\mathbf{H}_0)\mathbf{B}_0].
\label{eqA9}
\end{align} 
\indent By use of $\mathbf{D}\cdot\mathbf{k}_w=0$  from Eq.\ (\ref{eqA4}), $\mathbf{B}\cdot\mathbf{H}=\mathbf{E}\cdot\mathbf{D}$, and $\mathbf{a}\times(\mathbf{b}\times\mathbf{c})=(\mathbf{a}\cdot\mathbf{c})\mathbf{b}-(\mathbf{a}\cdot\mathbf{b})\mathbf{c}$, we have \\
\begin{equation}
\nabla\cdot\left[\check{\mathbf{I}}\frac{1}{2}(\mathbf{D}\cdot\mathbf{E}+\mathbf{B}\cdot\mathbf{H})\right]=2\cos\Psi\sin\Psi~ (\mathbf{D}_0\cdot\mathbf{E}_0)\mathbf{k}_w.
\label{eqA10}
\end{equation} \\ 
\indent Inserting Eqs.\ (\ref{eqA8}), (\ref{eqA9}), and (\ref{eqA10}) into Eq.\ (\ref{eqA1}), we find the left- and right-hand sides of Eq.\ (\ref{eqA1}) are equal.  Thus Eq.\ (\ref{eqA1}) is confirmed.
\\  \\

%%\section*{References}


\begin{thebibliography}{00}

\bibitem{r1}
I. Brevik and S. \r{A}. Ellingsen, ``Possibility of measuring the Abraham force using whispering gallery modes,'' Phys. Rev. A 81, 063830 (2010).

\bibitem{r2}
J. A. Stratton, Electromagnetic theory, (McGraw-Hill, NY, 1941), p. 159.

\bibitem{r3}
C. M\o ller, The theory of relativity, (Oxford University Press, London, 1952), p.205.

\bibitem{r4}
I. Brevik and S. \r{A}. Ellingsen, ``Detection of the Abraham force with a succession of short optical pulses,'' Phys. Rev. A 86, 025801 (2012).

\bibitem{r5}
A. Einstein, ``On the electrodynamics of moving bodies,'' Ann. Phys. Lpz. 17, 891 (1905).

\bibitem{r6}
I. Brevik, ``Comment on `Observation of a push force on the end face of a nanometer silica filament exerted by outgoing light','' Phys. Rev. Lett. 103, 219301 (2009).

\bibitem{r7}
I. Brevik and S. \r{A}. Ellingsen, ``Transverse radiation force in a tailored optical fiber,'' Phys. Rev. A (R) 81, 011806 (2010).

\bibitem{r8}
I. Brevik, ``Explanation for the transverse radiation force observed on a vertically hanging fiber,'' Phys. Rev. A 89, 025802 (2014).

\bibitem{r9}
C. Wang, ``Can the Abraham light momentum and energy in a medium constitute a Lorentz four-vector?'' Journal of Modern Physics 4, 1123 (2013); arXiv: 1409.4623.

\bibitem{r10}
T. Ramos, G. F. Rubilar, and Y. N. Obukhov, ``First principles approach to the Abraham-Minkowski controversy for the momentum of light in general linear non-dispersive media,'' J. Opt. 17, 025611 (2015).

\bibitem{r11}
C. Wang, ``Self-consistent theory for a plane wave in a moving medium and light-momentum criterion,'' accepted for publication in Canadian Journal of Physics; arXiv: 1409.5807.

\bibitem{r12}
I. Brevik, ``Experiments in phenomenological electrodynamics and the electromagnetic energy-momentum tensor,'' Phys. Rep. 52, 133 (1979), Sec. 1. 2.

\end{thebibliography}
\end{document}